\documentstyle[12pt,epsf]{article}

\newcommand{\plb}[3]{{{\it Phys.~Lett.}~{\bf B#1} (#3) #2}}
\newcommand{\npb}[3]{{{\it Nucl.~Phys.}~{\bf B#1} (#3) #2}}
\newcommand{\prd}[3]{{{\it Phys.~Rev.}~{\bf D#1} (#3) #2}}

\newcommand{\tinyfrac}[2]{\frac{\mbox{\footnotesize #1}}
{\mbox{\footnotesize #2}}}

\newcommand{\geqsim}{\,\raisebox{-0.6ex}{$\buildrel > \over \sim$}\,}
\newcommand{\be}{\begin{equation}}
\newcommand{\ee}{\end{equation}}
\newcommand{\ba}{\begin{eqnarray}}
\newcommand{\ea}{\end{eqnarray}}
\newcommand{\nn}{\nonumber}
\newcommand{\tl}[1]{\tilde{#1}}

\def\df{{\rm d}}

\def\hc{{\rm h.c.}}

\def\>{\mbox{\hspace{0.1cm}}}

\def\gev{\,{\rm GeV}}

\begin{document}
\thispagestyle{empty}
\begin{flushright}
RAL-TR-97-032\\ ULB-TH-97/14\\ {\tt hep-ph/9707436}
\end{flushright}
\vspace{5mm}
\begin{center}
{\Large \bf The Quasi-Fixed MSSM}\\
\vspace{15mm} {\large S.A.~Abel$^a$ {\small\it and} \/B.C.~Allanach$^b$}\\
\vspace{1cm}
$^a$ {\small\it Service de Physique Th\'eorique\\ 
Universit\'e Libre de Bruxelles\\
Bruxelles 1050, Belgium }\footnote{Address
after 1st Sept 1997: Theory Division, CERN CH-1211, Geneva 23}\\
\vspace{1cm}
$^b$ {\small\it Rutherford Appleton Laboratory 
\\Chilton, Didcot OX11 0QX, England}
\end{center} 
\vspace{2cm}

\begin{quote} 
\noindent The infra-red fixed points are determined for 
all the parameters of the MSSM. 
They dominate the renormalisation group running when 
the top-Yukawa is in the quasi-fixed point regime (i.e.\ large
at the GUT scale). We examine this behaviour analytically, 
by solving the full set of one-loop renormalisation
group equations in the approximation that the electroweak contributions are
negligible, and also numerically.
We find the 
quasi-fixed points for the top-quark trilinear couplings; 
$A_{U_{\alpha 3}} = A_{U_{3\alpha}}=-0.59 m_g$ independently of the 
input parameters at the unification scale. All the 
remaining parameters 
are significantly focused towards their true fixed points at the weak 
scale. We examine how this increases the predictivity of the MSSM in 
this regime.
\end{quote} 

\newpage

\section{Introduction}

The renormalisation group running of parameters from high 
to low scales is often dominated by infra-red stable fixed points.
Formally, these are points in parameter space where couplings
are invariant under the renormalisation group.
From any initial values at say the GUT scale, the couplings will flow
asymptotically toward the infra-red fixed points at lower scales.
In certain cases, a possible example being the top-quark Yukawa coupling,
the convergence is 
enough to increase the predictivity of the theory~\cite{irfps,graham}.
In the Minimal Supersymmetric Standard Model (MSSM) for instance,
in the limit that one can neglect $h_b,h_\tau $, the bottom and tau Yukawa
couplings,
a prediction for $h_t(m_t)$ yields a prediction for $\tan \beta$ through the
relation
\be
\sin \beta = \frac{m_t(m_t)}{v h_t(m_t)} \label{tanbpred}
\ee
where $m_t(m_t)\sim 167 \gev$ is the running top quark mass extracted from
experiment and $v=174.1 \gev$ is the Higgs vacuum expectation value parameter
extracted from $M_Z$.
The MSSM fixed point prediction for $h_t$ is 
$h_t(m_t)=\sqrt{7/18} g_3(m_t)$, where $g_3$ is the QCD gauge coupling.
For $m_t(m_t)\approx 167\gev$, $\alpha_s(m_t)=0.108$, we obtain the fixed
point prediction $h_t(m_t)=0.73$.
Substituting these values into Eq.(\ref{tanbpred}), however, yields
$\sin \beta = 1.32$, and hence this is an unphysical scenario. There are
then two possible scenarios within the MSSM:
\begin{itemize}
\item{The assumption of small $h_b,h_\tau$ is not valid and therefore $\tan
\beta$ is large, say $\tan \beta>30$. This possibility was discussed in
Ref.\cite{largetanb}.}
\item{The top quark Yukawa coupling is outside the domain of attraction of the
fixed point but $\tan \beta<30$ is small~\cite{graham}.}
\end{itemize}
In this paper, we shall examine what happens in the second scenario.
For larger values of $h_t(M_{GUT})$ a different type of behaviour
takes over 
which we shall refer to as quasi-fixed behaviour~(see 
\cite{carena,casas} and references therein). 
Large values of $h_t(M_{GUT})$ quickly converge to 
the envelope bounding the perturbative region, giving a
quasi-fixed point (QFP) prediction at $m_t$. The QFP limit is
formally defined as the landau pole of $h_t$ at the scale $M_{GUT}$.
Although the renormalised top coupling
is still running with scale $Q$ it is insensitive to $h_t(M_{GUT})$. 
We define
the quasi-fixed regime to be where $h_t(m_t)$ is close to 
the QFP prediction. 

\begin{figure}
\begin{center}
\epsfysize=6.21in
\epsfxsize=6.21in
\epsffile{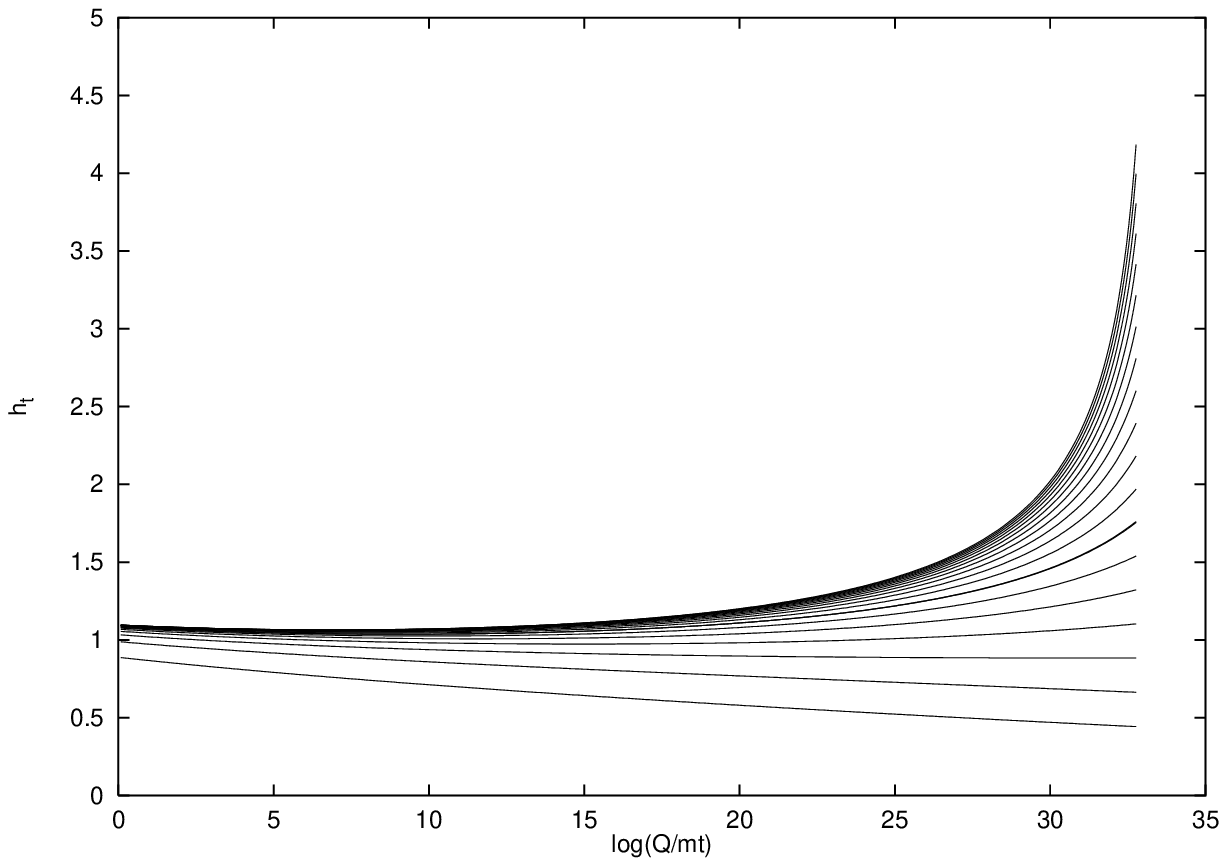}
\end{center}
\caption{The two-loop renormalisation of the top Yukawa coupling, $h_t$, for
$m_t=140\gev$. We have included electroweak and $h_b, h_\tau$ corrections and
$\tan \beta$ is determined seperately for each line by
Eq.(\protect\ref{tanbpred}).
The true fixed point is $h_t/g_3=0.9$ and is invariant under the
renormalisation group. 
Electroweak corrections make this higher than the naive value,
$ h_t=\sqrt{7/18}g_3 $. 
The QFP limit is formally defined as the $h_t$ trajectory for which $h_t$ has
a Landau pole at $M_{GUT}$.}
\label{hufig}
\end{figure}

These fixed and quasi-fixed behaviours (with electroweak 
and two-loop contributions included and for a top mass of 140~GeV) are shown
in figure (\ref{hufig}), where we show a numerical running for the MSSM, with 
a superpotential of the form  
\begin{equation}    
\label{super_mssm}                                           
W=h_U Q_L H_2 U_R + h_D Q_L H_1 D_R
 + h_E L H_1 E_R + \mu H_1 \varepsilon H_2,
\end{equation}    
in which generation indices are implied, and the superfields have the
standard definition~\cite{haberkane}. 
The program described in Ref.\cite{me,me2}
is the one we use to perform the running (without thresholds). 
We also take $\alpha_s(m_t)=0.108$ 
throughout, and define $M_{GUT}$ to be the scale $Q$ for which
$\frac{3}{5}\alpha_1(Q)=\alpha_2(Q)$. 
We do not unify $g_3$ with $g_2$ and $g_1$ because $g_3$ at the high scale is
sensitive to GUT threshold effects~\cite{thresh}.
The fixed point is only slightly dependent on the electroweak 
corrections and the focusing behaviour is not changed significantly by them,
although the numerical value of the QFP prediction is slightly modified. 
For this reason we shall neglect electroweak corrections in later analytic 
discussions of focusing behaviour and find that, when they are 
switched off, all our numerical and analytic results 
are in good agreement. From figure (\ref{hufig}), the quasi-fixed point 
prediction is $h_t(m_t)=1.09$, so that for $m_t=167$ GeV, we 
find $\tan \beta = 1.85$. 

This focusing property of $h_t$ is well known;
our aim in this paper is to examine what happens to the 
other couplings, in particular the soft supersymmetry breaking 
parameters, when $h_t$ is in the quasi-fixed regime. 
Many of the soft supersymmetry 
breaking terms also have QFPs; 
in the quasi-fixed regime, 
their low energy predictions are focused to running values which
are independent of the inputs at $M_{GUT}$. 
These soft parameters formally have true fixed point values as well,
but a necessary condition
for the true fixed point predictions to be valid is that $h_t$ is at its fixed
point prediction, which we have shown above to be untrue for small $\tan
\beta$ and so these values are not physically relevant. 
Some of the remaining couplings (which do not formally have QFPs) also have 
true fixed points, and they are significantly focused towards them.
They might then be said to exhibit quasi-fixed behaviour even though they 
are not fully independent of the inputs at the GUT scale. 

The parameter that 
always has a QFP 
and that shows the  most striking convergence is $A_t$, the
trilinear stop coupling~\cite{carena}.
Its running is shown in figure (2) for various starting values 
at the GUT scale. When $h_t$ is in the
quasi-fixed regime, this coupling is determined to be,
\be
\label{at}
A_t (m_t) = -0.59\mbox{~}m_{g}.
\ee  
(We use $A_t$ as shorthand for $A_{U_{33}}$ and $m_{g}$ for the gluino
mass). In this and subsequent occasions where we 
demonstrate numerically the (lack of) dependence 
on the initial conditions, we take, 
simply for convenience, the supersymmetry breaking terms,
\begin{eqnarray}                                               
-\delta {\cal{L}}&=&m^2_{ij}z_i z_j^*  + \tinyfrac{1}{2}M_A
\lambda_A \lambda_A \nonumber\\ 
&& + A_U h_U{\tl q}^*_L h_2 {\tl u}_R 
+ A_D h_D{\tl q}^*_L h_1 {\tl d}_R
+ A_E h_E{\tl l}^* h_1 {\tl e}_R + B \mu h_1 \varepsilon h_2  +\hc,
\label{lag}
\end{eqnarray}
to be of the `constrained' MSSM (CMSSM) form at the 
GUT scale; 
\ba
\label{degen}
A_{U,D,E}&=&A \nonumber\\
m^2_{ij}   &=&\delta_{ij} m_0^2\nonumber\\
M_A      &=&m_{1/2}.\label{cmssm}
\ea
In Eq.(\ref{lag}), $z_i$ stand for all scalar fields in the theory. We
will specifically refer to
$m^2_{Q_{ij}}$, $m^2_{U_{ij}}$, $m^2_{D_{ij}}$ for the left-handed
squarks, the right-handed up and the right-handed down squark soft mass
parameters respectively.

\begin{figure}
\begin{center}
\epsfysize=6.21in
\epsfxsize=6.21in
\epsffile{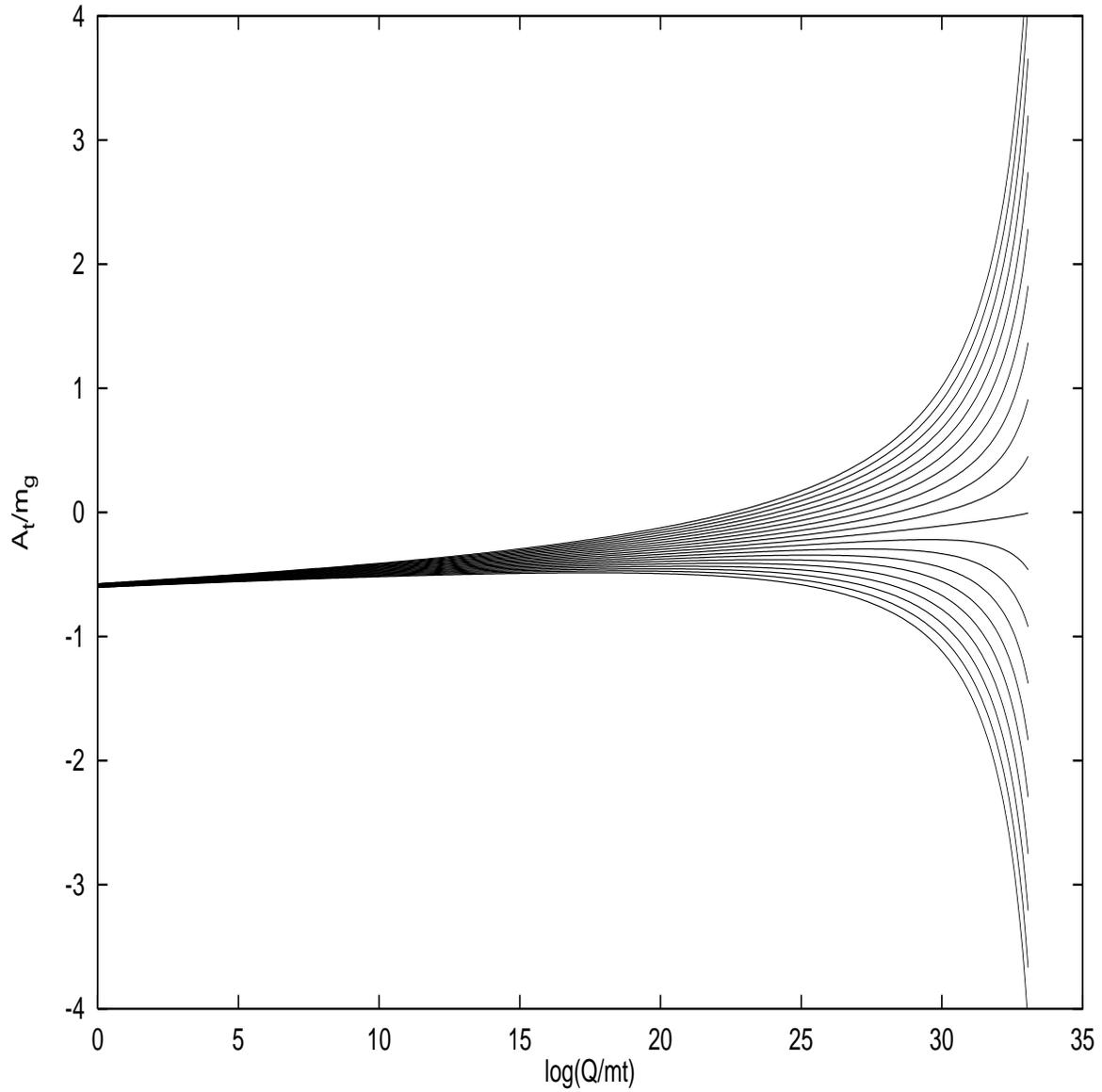}
\end{center}
\caption{Two-loop renormalisation of $A_t/m_g$ in the CMSSM
for various different 
initial values, with $h_t (GUT)=5 g_3(GUT)$ and $m_t=175$\gev. 
All electroweak and Yukawa contributions are included.}
\label{atfig} 
\end{figure}

Later we shall see why the strong focusing seen in figure (\ref{atfig})
is quite independent of the pattern of supersymmetry breaking. Hence, 
since both $A_t$ and $\tan\beta $ are fully determined in the
quasi-fixed regime, there are at least two less free parameters in
{\em all} versions of the MSSM (except trivially the no-scale models). 
The mass spectrum of the CMSSM, for example, depends 
(to first order in $m_b/m_t$) only on $m_0$ and $m_{1/2}$~\cite{carena}.

The behaviour of $A_t$ suggests that other low scale soft parameters 
may be quite insensitive to the high scale inputs and, 
as stated above, we shall indeed find this to be the case,
by solving the renormalisation group equations (RGEs)
analytically for all the soft supersymmetry breaking terms 
including the flavour changing ones. In the CMSSM we shall show that
combinations of the latter also have QFPs. 

\section{The fixed points}

First, let us derive the fixed point values that the couplings would run
towards in
the infra-red if the top-quark fixed point were valid. They correspond
to the point where the beta function vanishes for the coupling in 
question~\cite{irfps,graham}.   
So, for example, using the RGEs of 
Refs.\cite{rges} without electroweak or two-loop contributions,
\begin{eqnarray}
\frac{\df g_3}{\df t}& =& -3 \frac{ g_3^3}{16\pi^2}\nonumber \\ 
\frac{\df h_t}{\df t}& =& \frac{h_t}{16\pi^2}(6 h_t^2
-\tinyfrac{16}{3}g_3^2 ),
\end{eqnarray} 
the RGE of $h_t^2 /g_3^2 $~\cite{graham} vanishes when 
\be
\frac{h_t^2}{g_3^2}=\frac{7}{18}.
\ee
In the evolution from from high to low energy 
scales, $h^2_t/g^2_3$ runs towards 
this value because the fixed point is infra-red stable.
A similar procedure can be carried out for all of the 
couplings in the MSSM, again neglecting electroweak corrections
and all but the top Yukawa contributions to the running. 
If there are any large hierarchies in the soft masses (comparable in size to
the hierarchy between the top and bottom Yukawa couplings) then this
approximation may not always be valid. We will investigate this possibility in
detail elsewhere~\cite{ourstobe}.
Normalising the couplings by the 
gluino mass, so that $\tl{A}_t\equiv A_t/m_g $, $\tl{B}_t\equiv B/m_g $
and $\tl{m}_i^2 \equiv m_i^2/m_g^2$,
we find the following infra-red fixed points;
\ba
-\tl{A}_U = \left( \begin{array}{ccc}
\frac{25}{18} & \frac{25}{18} & 1 \\
\frac{25}{18} & \frac{25}{18} & 1 \\
1 & 1 & 1 \\ \end{array} \right) && 
-\tl{A}_D = \left( \begin{array}{ccc}
\frac{16}{9} & \frac{16}{9} & \frac{16}{9} \\
\frac{16}{9} & \frac{16}{9} & \frac{16}{9} \\
\frac{89}{54} & \frac{89}{54} & \frac{89}{54} \\ \end{array} \right) \nn \\
\tl{m}^2_{Q} = \left( \begin{array}{ccc}
\frac{8}{9} & 0 & 0 \\
0 & \frac{8}{9} & 0 \\
0 & 0 & \frac{41}{54} \\ \end{array} \right) &&
\tl{m}^2_{U} = \left( \begin{array}{ccc}
\frac{8}{9} & 0 & 0 \\
0 & \frac{8}{9} & 0 \\
0 & 0 & \frac{17}{27} \\ \end{array} \right)  \nn \\
\tl{m}^2_{D} = \left( \begin{array}{ccc}
\frac{8}{9} & 0 & 0 \\
0 & \frac{8}{9} & 0 \\
0 & 0 & \frac{8}{9} \\ \end{array} \right) &&
\begin{array}{c}
\tl{m}^2_{2} =  - \frac{7}{18} \\
\mu^2 = 0\\
\tl{B} = \frac{7}{18}.\\
\end{array} \label{FPs}
\ea
The parameter $m^2_{2}$ has a negative fixed point, because it is 
dragged down by the top Yukawa coupling. This 
guarantees electroweak symmetry breaking in the fixed regime, 
and underlies the breaking that is found away from it. 
Fixed points for a subset of Eq.(\ref{FPs})
($\tl{A}_t$, $\mu$, $\tl{\mu_2}^2$, $\tl{m}_{t}^2$, $\tl{B}$,
$\tl{m}_{Q_t}^2$) were previously found by Lanzagorta and 
Ross~\cite{graham}, and our
results agree with those.
Notice that the fixed points are independent of 
the quark basis in which we choose to calculate them (so long as 
it is still accurate to assume $h_t \approx h_{U_{33}} $ to be the 
dominant Yukawa coupling in that basis). 
(Note, however, that if we choose a basis where
the down quarks' masses are diagonal at $m_t$, then $A_{D_{i\neq j}}$
is of course not defined at $ m_t$ 
whereas $A_{U_{i\neq j}}$ is, and vice versa.) 

\section{Exact one-loop RGE solutions}

There is obviously a difference between the fixed point 
value of $A_t$, and that in Eq.(\ref{at}). This is because 
Eq.(\ref{at}) is in fact also a quasi-fixed point; even though 
it is focused, $A_t$ is still running at $m_t$. 
Evidently we need to solve the
renormalisation group equations and apply the solutions
to the quasi-fixed regime.
In this section we do this analytically
and exactly to one loop and in the approximation that the 
electroweak and non-$h_t$ 
Yukawa contributions are negligible. 
It is simplest to do this in terms 
of the ratio of strong coupling constants; 
\be
r{\scriptstyle (Q)} \equiv \frac{\alpha_3{\scriptstyle (M_{GUT}})}
{\alpha_3{\scriptstyle (Q)}}=
1-\mbox{\footnotesize 6}\>
{\alpha_3{\scriptstyle (M_{GUT}) }}\log {\scriptstyle (
\frac{Q}{M_{GUT}})}. \nn \\
\ee
The RGE for $R\equiv h_t^2/g_3^2 $ may be solved~\cite{graham} such that
\be
R {\scriptstyle (r) }=\frac{7}{18-18\> r^{7/9} + 7\> r^{7/9} R_0^{-1} }
\ee
where $R_0\equiv R {\scriptstyle (M_{GUT}) }$. The fixed point corresponds
to 
$r\rightarrow 0$, and the QFP to $R_0\rightarrow \infty
$ with $r=1/(25 \times 0.108)=0.37$, so that $R(r)<R^{QFP}(r)=0.72$.
Next we solve for following combinations of couplings to 
find solutions in terms of $\tl{A}_t$ and $\tl{m}^2_{U_{33}} $; here the
suffix $0$ again implies values taken at the GUT scale and $\alpha,
\beta=1,2,3$ and $i,j=1,2$
\ba
\label{solutions}
\tl{B}-\tinyfrac{1}{2} \tl{A}_t &=& -\tinyfrac{8}{9} (r-1) + 
r (\tl{B}-\tinyfrac{1}{2}\tl{A}_t)|_0 \nn\\
&&\nn\\
\tl{A}_{D_{3 \alpha }}-\tinyfrac{1}{6}\tl{A}_t &=& 
\tinyfrac{40}{27} (r-1) + r (\tl{A}_{D_{3 \alpha }}-\tinyfrac{1}{6}\tl{A}_t)|_0
\nn\\
&&\nn\\
\tl{A}_{D_{i\alpha}} &=& \tinyfrac{16}{9} (r-1) + r \> \tl{A}_{D_{i\alpha }}|_0
\nn\\
&&\nn\\
\tl{A}_{U_{ij}}-\tinyfrac{1}{2}\tl{A}_t &=& 
\tinyfrac{8}{9} (r-1) + r (\tl{A}_{U_{ij}}-\tinyfrac{1}{2}\tl{A}_t)|_0 \nn\\
&&\nn\\
\tl{A}_{U_{3i}}-\tl{A}_t &=&  r \left(\frac{R\> r^{7/9}}{R_0}\right)^{1/6} 
(\tl{A}_{U_{3i}}-\tl{A}_t)|_0 \nn\\
&&\nn\\
\tl{A}_{U_{i3}}-\tl{A}_t &=&  r \left(\frac{R\> r^{7/9}}{R_0}\right)^{1/12} 
(\tl{A}_{U_{i3}}-\tl{A}_t)|_0 \nn\\
&&\nn\\
\tl{m}^2_1& = &r^2 \> \tl{m}_1^2|_0\nn\\
&&\nn\\
\tl{m}^2_2-\tinyfrac{3}{2} \tl{m}^2_{U_{33}} &=& -\tinyfrac{4}{3}(1-r^2) + r^2 (
\tl{m}^2_2-\tinyfrac{3}{2} \tl{m}^2_{U_{33}})|_0 \nn\\
&&\nn\\
\tl{m}^2_{Q_{33}}-\tinyfrac{1}{2} \tl{m}^2_{U_{33}} &=& \tinyfrac{4}{9}(1-r^2) + r^2 (
\tl{m}^2_{Q_{33}}-\tinyfrac{1}{2} \tl{m}^2_{U_{33}})|_0 \nn\\
&&\nn\\
\tl{m}^2_{Q_{ii}} &=& \tinyfrac{8}{9}(1-r^2) + r^2 \> \tl{m}^2_{Q_{ii}}|_0  \nn\\
&&\nn\\
\tl{m}^2_{U_{ii}} &=& \tinyfrac{8}{9}(1-r^2) + r^2
\> \tl{m}^2_{U_{ii}}|_0  \nn\\
&&\nn\\
\tl{m}^2_{D_{\alpha\alpha}} &=& \tinyfrac{8}{9}(1-r^2) + r^2 \> 
\tl{m}^2_{D_{\alpha\alpha}}|_0  \nn\\
&&\nn\\
\tl{m}^2_{Q_{\alpha\beta}} &=&  r^2 \left(\frac{R\> r^{7/9}}{R_0}\right)^{1/12} 
\tl{m}^2_{Q_{\alpha\beta}}|_0  \mbox{\hspace{1cm}$\alpha\beta = i3$ or $3i
$}\nn\\
&&\nn\\
\tl{m}^2_{U_{\alpha\beta}} &=&  r^2 \left(\frac{R\> r^{7/9}}{R_0}\right)^{1/6} 
\tl{m}^2_{U_{\alpha\beta}}|_0  \mbox{\hspace{1cm}$\alpha\beta = i3$ or $3i
$},  
\ea
with the other off-diagonal squark mass squareds $\tl{m}_o^2$ being
\be
\label{mo}
\tl{m}_o^2 = r^2 \tl{m}_o^2|_0. \label{offdiag}
\ee
We stress that these solutions are for generic couplings\footnote{Given the
proviso of no large hierarchies in soft masses.}, and care should
be taken over the choice of basis especially for the
$\tl{A}$-terms. One instance is the basis
used in Ref.\cite{me2}, where the down quarks are diagonal, and the 
up quark Yukawas are diagonalised just by a rotation by the CKM matrix on the
left
handed up quarks. In this basis, $\tl{A}_{3i}$ behaves more like
$\tl{A}_{ij}$ because of the cancellation of certain terms in the RGE\@. 
Now we need to solve for $\tl{A}_t$ and $\tl{m}^2_{U_{33}}$ themselves; 
the RGE for $\tl{A}_t $ becomes    
\be
r \frac{\df \tl{A}_t}{\df r} =  \tinyfrac{16}{9}+2 \tl{A}_t R + \tl{A}_t 
\ee
and gives
\ba
\label{atsolution}
\tl{A}_t & = & \tinyfrac{18}{7} R \left(
-1+r^{7/9}(\tinyfrac{16}{9}-\tinyfrac{7}{9} r) \right) + \nn\\
&&\hspace{3cm} \left(\frac{R}{R_0}\right) r^{7/9} \left( r
\tl{A}_t|_0 -\tinyfrac{16}{9}(1-r)  \right).
\ea
To solve for $\tl{m}^2_{U_{33}}$, we form the RGE for
$Z \equiv \tl{m}^2_{U_{33}}+
\tl{m}^2_{Q_{33}}+\tl{m}_2^2-\tl{A}_t^2$,
\be
r \frac{\df Z}{\df r} =  -\tinyfrac{32}{9} -\tinyfrac{32}{9} \tl{A}_t + 2
Z R + 2 Z, 
\ee
whose solution is
\ba
\label{Zeq}
Z & = &  \tinyfrac{32}{9} R \> r^{7/9}(1-r)^2 + \nn\\
&& \hspace{1cm} \tinyfrac{1}{9}\left(\frac{R}{R_0}\right) r^{7/9}
\left( 32 \> r(1-r)\tl{A}_t|_0 + 9\> r^2 Z|_0 - \tinyfrac{16}{9}(7-25 r)(1-r) \right).
\ea
Substituting from Eq.(\ref{solutions}) into $Z$ gives,
\be
\label{mtrsolution}
\tl{m}^2_{U_{33}}-\tinyfrac{1}{3}\tl{A}_t^2-\tinyfrac{1}{3}Z = 
\tinyfrac{8}{27} (1-r^2) + \tinyfrac{1}{3} r^2
(2  \tl{m}^2_{U_{33}} -\tl{m}^2_{Q_{33}}-\tl{m}^2_2 )|_0.
\ee
Eqs.(\ref{solutions},\ref{mo},\ref{atsolution},\ref{Zeq},\ref{mtrsolution})
are completely general solutions to the full set of one-loop 
RGEs when $h_t$ and $g_3$ are dominant. These solutions are consistent with
Ref.\cite{brax}, where electroweak corrections were 
also included. We have presented them here in a form which makes 
the dependence on $\tl{A}_t$ and $Z$, two parameters which always 
have QFPs, clear, and makes the existence of additional QFPs 
obvious in more constrained models. 
(Here we omit electroweak corrections 
principally because they only serve to obscure the fixed point
structure which is governed by $h_t$ and $g_3$.) We also prefer to leave the 
dependence on $R_0$ explicit, rather than substitute $R^{QFP}$.

We are now able to make some comments more specific to the MSSM 
in the quasi-fixed regime. 
Firstly note that, consistent with 
Ref.\cite{carena}, 
at the QFP ($R_0^{-1}=0 $), both $\tl{A}_t$ and $Z$ are independent 
of {\em any} \/of the input parameters. This then is the reason for the 
remarkable convergence of $\tl{A}_t$ seen in figure (\ref{atfig}) and is 
in fact true of $\tl{A}_{U_{\alpha 3}}$ and $\tl{A}_{U_{3 \alpha}}$
although, generally, $\tl{A}_t$ converges most strongly. Because their
leading terms are dependent only on $r$, it makes sense to refer
to them as QFPs as well. 
Inserting $R=0.72$ and $r=0.37$ into Eq.(\ref{atsolution}), we find the 
quasi-fixed value for $\tl{A}_{U_{\alpha 3}}$ and
$\tl{A}_{U_{3 \alpha}}$ consistent with the numerical
determination in Eq.(\ref{at});
\be
\tl{A}_{U_{\alpha 3}} =
\tl{A}_{U_{3 \alpha}} = -0.58.
\ee
The remaining $A$ parameters converge approximately linearly to their fixed
point, 
and the $m^2$ terms quadratically to theirs. This dependence is 
due to the diverging gluino mass
$m_{g}= r^{-1}m_{g}|_0$.

\begin{figure}
\begin{center}
\hspace*{-0.5in}
\epsfysize=6.21in
\epsfxsize=6.21in
\epsffile{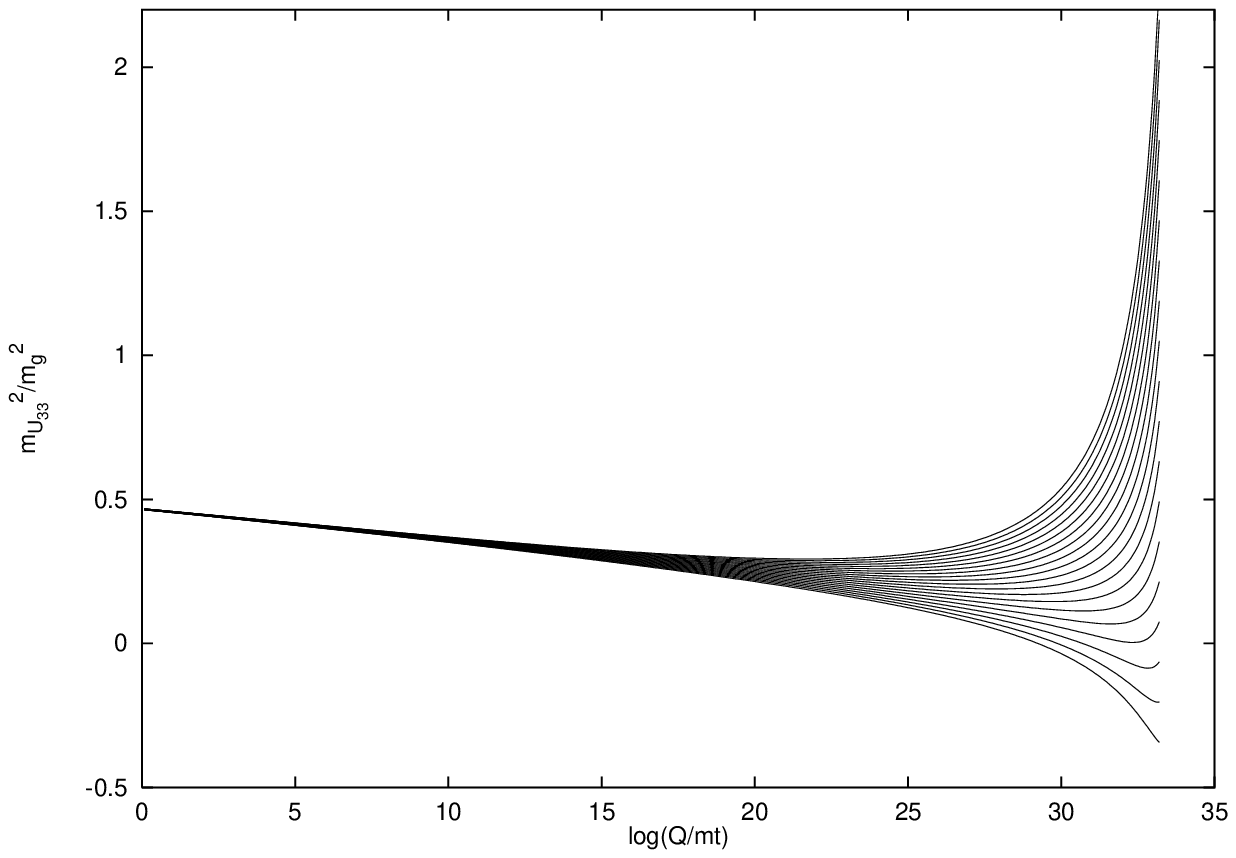}
\end{center}
\caption{Two-loop renormalisation of $m^2_{{U}_{33}}/m_g^2$ in the 
CMSSM, with $h_t (GUT)=5 g_3(GUT)$ and $m_t=175$\gev.
All electroweak and Yukawa contributions are included.}
\label{mtrfig} 
\end{figure}

\begin{figure}
\begin{center}
\epsfysize=6.21in
\epsfxsize=6.21in
\epsffile{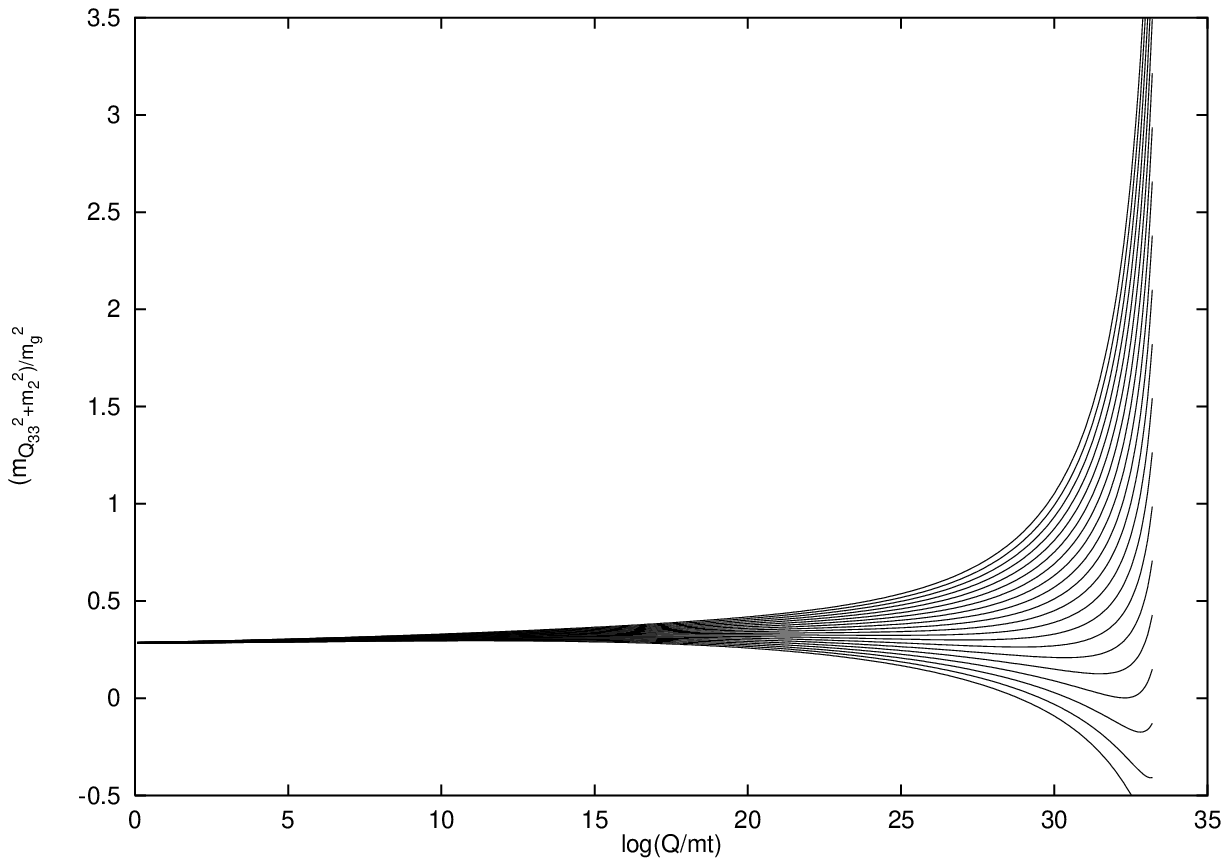}
\end{center}
\caption{Two-loop renormalisation of
$(m^2_{Q_{33}}+ m^2_2)/m_g^2$ in the CMSSM, with $h_t (GUT)=5
g_3(GUT)$ and $m_t=175$\gev. All electroweak and Yukawa 
contributions are included.}
\label{combfig1} 
\end{figure}

\begin{figure}
\begin{center}
\epsfysize=6.21in
\epsfxsize=6.21in
\epsffile{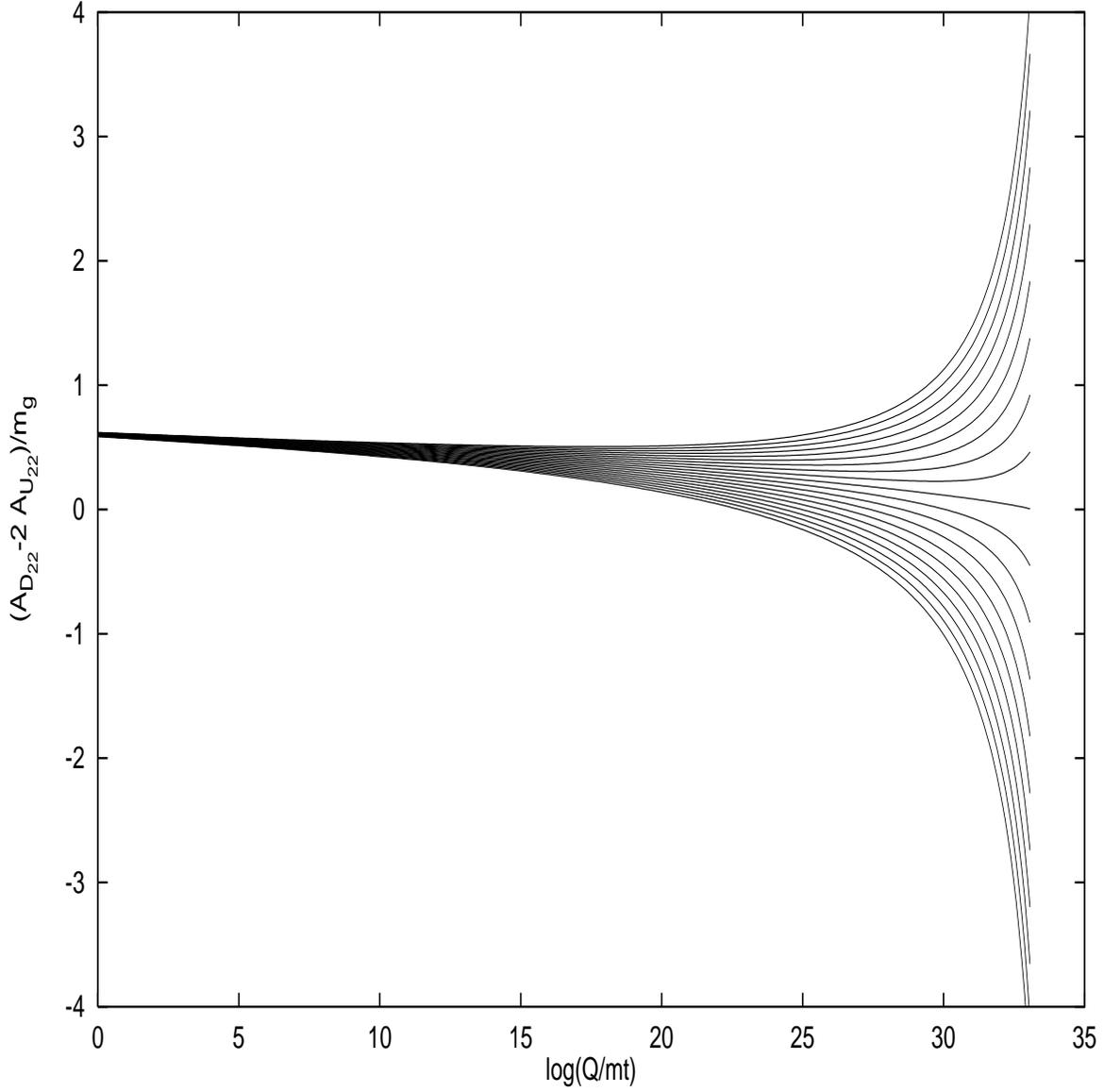}
\end{center}
\caption{Two-loop renormalisation of
$(\tl{A}_{D_{22}}-2 \tl{A}_{U_{12}})/m_g$ in the CMSSM, 
with $h_t (GUT)=5 g_3(GUT)$ and $m_t=175$\gev in the CMSSM\@. All electroweak 
and Yukawa contributions are included.}
\label{combfig2} 
\end{figure}

Coincidentally, in the constrained MSSM (CMSSM), there are a number 
of other QFPs which can be deduced from the solutions above
(two of which were noted in Ref.\cite{carena})
including one for $\tl{m}^2_{U_{33}}$ itself
(which is why we wrote Eq.(\ref{solutions}) in terms of it). 
In our analytic approximation these are 
\ba
\tl{A}_{U_{\alpha 3}} =
\tl{A}_{U_{3 \alpha}} &=& -0.58 \nn\\
\tinyfrac{6}{5}\tl{A}_{D_{3 \alpha}}-2 \tl{A}_{U_{i j}}&=& 0.46 \nn\\
\tl{A}_{D_{i \alpha}}-2 \tl{A}_{U_{i j}}&=&0.58 \nn\\
\tl{m}^2_{U_{33}} &=&0.53 \nn\\
\tl{m}^2_{Q_{33}}+\tl{m}_2^2 &=&0.28 \nn\\
\tl{m}^2_{Q_{ii}}+2 \tl{m}_2^2 &=&0.04 \nn\\
\tl{m}^2_{U_{ii}}+2 \tl{m}_2^2 &=&0.04 \nn\\
\tl{m}^2_{D_{\alpha\alpha}}+2 \tl{m}_2^2 &=&0.04 \nn\\
\tl{m}^2_1+2 \tl{m}_2^2 &=&-0.73 \nn\\
\ea
where the meanings of the indices are as before. 
In the CMSSM, all of these combinations are independent of the 
GUT scale parameters
when $R_0\rightarrow\infty $, and converge to their QFPs roughly in the same
manner  
as $\tl{A}_t$. The value of $R_0$ regulates how quickly 
the convergence to the QFP occurs (with $R_0^{-1}=0$ giving instant 
convergence).
The running of $\tl{m}^2_{U_{33}}$ for various starting values 
in the CMSSM is shown in figure (\ref{mtrfig}), 
and the running of the combinations $\tl{m}^2_{Q_{33}}+\tl{m}_2^2$
and $ \tl{A}_{D_{22}}-2 \tl{A}_{U_{22}} $ are shown, 
with electroweak and two loop contributions included,
in figures (\ref{combfig1}) and (\ref{combfig2}) respectively. 
These QFPs allow us to assess the net effect of electroweak 
and two-loop contributions; numerically we find 
\ba
\tl{A}_{U_{\alpha 3}} =
\tl{A}_{U_{3 \alpha}} &=& -0.59 \nn\\
\tinyfrac{6}{5}\tl{A}_{D_{3 \alpha}}-2 \tl{A}_{U_{i j}}&=& 0.49 \nn\\
\tl{A}_{D_{i \alpha}}-2 \tl{A}_{U_{i j}}&=&0.60 \nn\\
\tl{m}^2_{U_{33}} &=&0.47 \nn\\
\tl{m}^2_{Q_{33}}+\tl{m}_2^2 &=&0.29 \nn\\
\tl{m}^2_{Q_{ii}}+2\tl{m}_2^2 &=&0.04 \nn\\
\tl{m}^2_{U_{ii}}+2\tl{m}_2^2 &=&-0.01 \nn\\
\tl{m}^2_{D_{\alpha\alpha}}+2\tl{m}_2^2 &=&-0.02 \nn\\
\tl{m}_1^2+2 \tl{m}^2_2 &=&-0.72 \nn\\
\ea
which agree well with the analytic approximation. We stress 
that the analytic and numerical results provide a powerful 
(and successful) cross check. 

These fixed points reveal a remarkable fact for the CMSSM\@. 
We have already noted that $m^2_2$ is driven negative, a requirement 
of electroweak symmetry breaking. These QFPs tell us that, 
in the quasi-fixed regime, all the remaining squark mass-squareds 
are {\em guaranteed to be positive}\/ at low energy scales, 
even if their GUT scale values are zero. 

Finally in this section, when can we say that 
we are in the quasi-fixed regime?
As a working definition, let us define the quasi-fixed regime 
to be when $\tl{A}_t$ varies at the weak scale by 
less than 5\% of its deviation at the GUT scale. Using 
the solution for $\tl{A}_t$ with $r=0.37$, we find that 
this is true for a remarkably low value; 
$R_0\geqsim 2.5 $ or 
\be
h_t(M_{GUT})\geqsim 1.1
\ee
This is the case for example in models which, with string 
theory in mind, take as their Yukawa couplings 
$h_t=\sqrt{2} g \geqsim 1.2$ (assuming that renormalisation 
between the Planck and GUT scales doesn't reduce the Yukawa 
couplings significantly.)

\section{The Spectrum}

As well as these QFPs for the specific case of the CMSSM, our 
solutions quite generally imply a convergence 
towards the true fixed points in the 
infra-red. 
In this section, we return to non-specific 
patterns of supersymmetry breaking, to consider what this 
convergence means for the MSSM in general. 
We will call the focusing effect quasi-fixed behaviour.
 
Clearly quasi-fixed behaviour indicates that uncertainties in our 
knowledge of GUT (or Planck) scale physics are less important 
at low energy. To make this statement quantitative, we introduce a `focal
factor' $F(Y)$ which is defined for some
quantity $Y(m_t)$ by the deviation in $Y(m_t)$ produced by independently
varying all GUT
scale input parameters by 100\% of $m_{g}$ (each deviation being added in
quadrature).
This value is dependent upon how near one is to the top Yukawa QFP, and here
we assign $h_t(M_{GUT})=2$ as an example. $F(Y)$ then gives a measure
of the dependence of each parameter upon initial conditions\footnote{And is
also equivalent to the 1$\sigma$ error in the prediction for $Y(m_t)$
produced by assuming that the GUT scale parameters each have a 1$\sigma$
deviation of $m_{\tl{g}}$ around $m_{\tl{g}}$.}.

Our prejudice is that the soft supersymmetry breaking terms should all 
be at least of the same order of magnitude. Accordingly, we choose 
to determine the spectrum around the central values at the GUT scale;
\be
\tl{m}_{ij}^2=\tl{A}=\tl{B}=1
\ee
In our analytic approximation we find
\ba
-\tilde{A}_U = \left( \begin{array}{ccc}
0.67 & 0.67 & 0.59 \\
0.67 & 0.67 & 0.59 \\
0.59 & 0.59 & 0.59 \\ \end{array} \right) &&
-\tilde{A}_D = \left( \begin{array}{ccc}
0.75 & 0.75 & 0.75 \\
0.75 & 0.75 & 0.75 \\
0.72 & 0.72 & 0.72 \\ \end{array} \right) \nn \\
\tilde{m}^2_{U} = \left( \begin{array}{ccc}
0.90 & 0.14 & 0.08 \\
0.14 & 0.90 & 0.08 \\
0.08 & 0.08 & 0.54 \\ \end{array} \right) &&
\tilde{m}^2_{D} = \left( \begin{array}{ccc}
0.90 & 0.14 & 0.14 \\
0.14 & 0.90 & 0.14 \\
0.14 & 0.14 & 0.90 \\ \end{array} \right) \nn \\
\tilde{m}^2_{Q} = \left( \begin{array}{ccc}
0.90 & 0.14 & 0.10 \\
0.14 & 0.90 & 0.10 \\
0.10 & 0.10 & 0.72 \\ \end{array} \right) &&
\begin{array}{c}
\tilde{m}_2^2 = -0.41 \\
\end{array}
\label{qfpsol}
\ea
and the numerical values when all electroweak and 
two-loop corrections are included are 
\ba
-\tilde{A}_U = \left( \begin{array}{ccc}
0.82 & 0.87 & 0.59 \\
0.82 & 0.97 & 0.59 \\
0.59 & 0.59 & 0.59 \\ \end{array} \right) &&
-\tilde{A}_D = \left( \begin{array}{ccc}
0.96 & 0.96 & 0.96 \\
0.96 & 0.96 & 0.96 \\
0.90 & 0.90 & 0.90 \\ \end{array} \right) \nn \\
\tilde{m}^2_{U} = \left( \begin{array}{ccc}
0.91 & 0.14 & 0.07 \\
0.14 & 0.91 & 0.06 \\
0.07 & 0.06 & 0.48 \\ \end{array} \right) &&
\tilde{m}^2_{D} = \left( \begin{array}{ccc}
0.91 & 0.14 & 0.14 \\
0.14 & 0.91 & 0.14 \\
0.14 & 0.14 & 0.91 \\ \end{array} \right) \nn \\
\tilde{m}^2_{Q} = \left( \begin{array}{ccc}
0.98 & 0.14 & 0.10 \\
0.14 & 0.98 & 0.11 \\
0.10 & 0.11 & 0.75 \\ \end{array} \right) &&
\begin{array}{c}
\tilde{m}_2^2 = -0.46 \\
\end{array} 
\label{connumer}
\ea
Since $\tl{B}$ and $\tilde{\mu}$ are determined by the minimisation of the 
effective potential they are no longer free parameters. 
The focal factors are
\ba
F(\tilde{A}_U) = \left( \begin{array}{ccc}
.19(.41) &.19(.41)  &.02(.39)  \\
.19(.41) &.19(.41)  &.02(.39)  \\
.02(.30) &.02(.30)  &.02(.02)  \\ \end{array} \right) &&
F(\tilde{A}_D) = \left( \begin{array}{ccc}
.37(.37) & .37(.37) & .37(.37) \\
.37(.37) & .37(.37) & .37(.37) \\
.31(.37) &.31(.37)  &.31(.37)  \\ \end{array} \right) \nn \\
F(\tilde{m}^2_{U}) = \left( \begin{array}{ccc}
.14(.14) & .14(.14) &.08(.08)  \\
.14(.14) &.14(.14)  &.08(.08)  \\
.08(.08) &.08(.08)  &.02(.17) \\ \end{array} \right) &&
F(\tilde{m}^2_{D}) = \left( \begin{array}{ccc}
.14(.14) &.14(.14)  &.14(.14)  \\
.14(.14) & .14(.14) &.14(.14)  \\
.14(.14) & .14(.14) &.14(.14)  \\ \end{array} \right) \nn \\
F(\tilde{m}^2_{Q}) = \left( \begin{array}{ccc}
.14(.14) &.14(.14)  &.11(.11)  \\
.14(.14) &.14(.14)  &.11(.11)  \\
.11(.11) &.11(.11)  &.07(.11)\\ \end{array} \right) &&
\begin{array}{c}
F(\tl{m}^2_2)=.08(.11) \\
\end{array} \label{Fs}
\ea
where the un-bracketed entries are valid in a more constrained 
(but still non-universal) case where
$m_{ij}^2 (M_{GUT})=m_0^2$, $A_{U,D,E}(M_{GUT})=A$ etc.
and the bracketed entries
are true for the general (i.e.\ unconstrained) MSSM. 

\begin{figure}
\begin{center}
\epsfysize=6.21in
\epsfxsize=6.21in
\epsffile{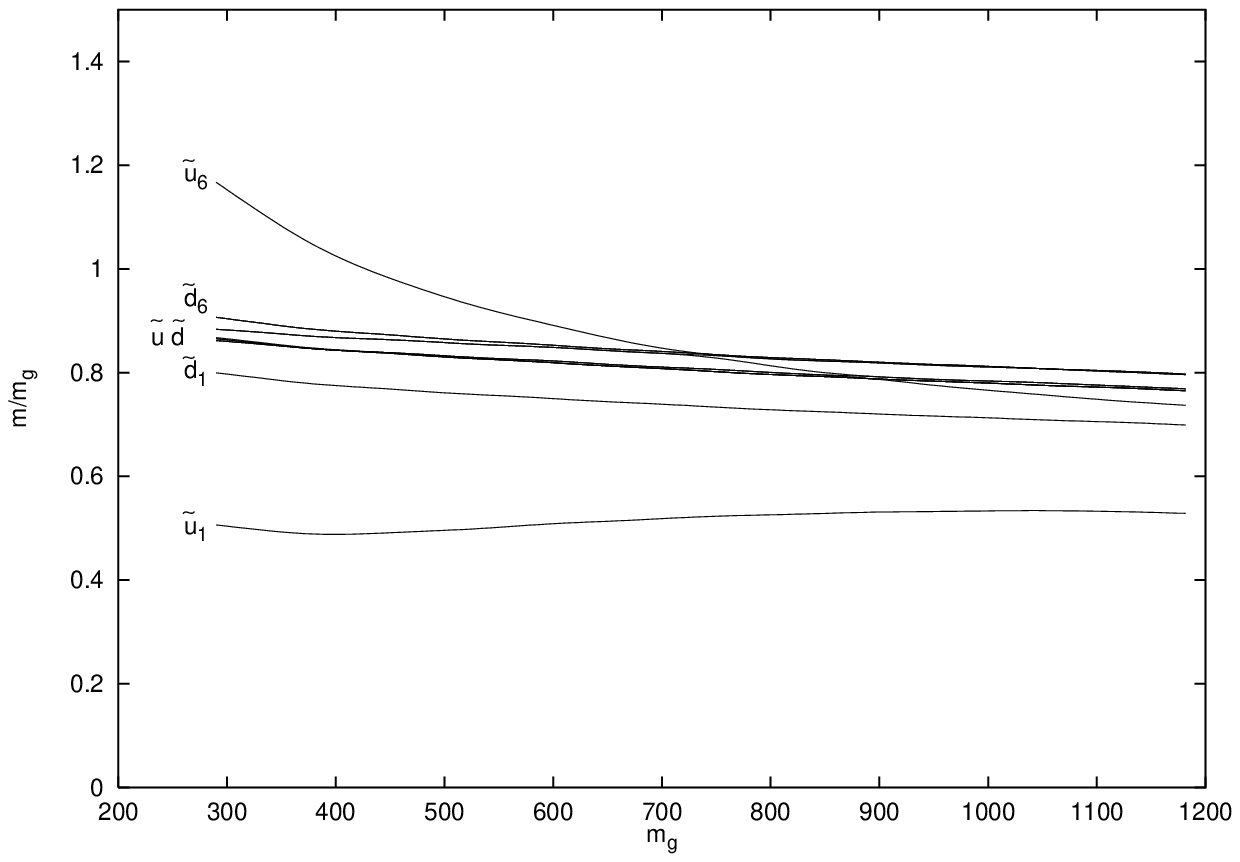}
\end{center}
\caption{Mass spectrum of the squarks when 
$h_t (GUT)=2$ and $m_t=175$\gev as 
a function of $m_g$ in the
MSSM with `central values'. 
All electroweak and Yukawa contributions are included. 
An error of $\pm m_g$ at the GUT scale, corresponds to 
$\pm 0.18$ in this spectrum.}
\label{spectrum} 
\end{figure}

As an example, let us determine the squark mass-spectrum for these 
central values, whose values we expect to be close to the real ones
(modulo the respective focusing factors). 
The mass spectrum depends (to first order) only 
on $A_t$ through the mixing in the top squark mass matrix. 
Since $\tl{A}_t$ always has a QFP, $\tan \beta$ is 
determined, and $\mu$ and $B$ are determined by 
minimising the effective potential, 
the entire spectrum depends only on $\tl{m}_{ij}^2 $ and $m_g$.
Furthermore, the dependence on $\tl{m}_{ij}^2 $ is also reduced by 
the infra-red focusing factors above. 
The minimisation to find $\mu $ and $B$ 
may be done in the same manner as in described in
Refs.\cite{barger,me}, 
and yields the mass
spectrum for the squarks shown in figure (\ref{spectrum}), which 
was determined for
negative $\mu $. The masses are almost proportional to 
$m_g$ (deviation from proportionality coming from $M^2_Z$ and $m_t^2$
terms in the mass-squared matrices). 

We then expect various models of supersymmetry breaking to give 
a spectrum close to this one in the quasi-fixed regime 
(modulo the `uncertainty' represented by the focus factors above). 
Even allowing for an uncertainty of 100 \% in our knowledge 
of the pattern of supersymmetry breaking at the GUT scale, the 
focal factors tell us that the mass squareds are determined at the 
weak scale to better than 14\% of $m_g^2$.
However, note that the off-diagonal elements of the 
squark mass-squareds $\tl{m}_o^2$ are given by 
$\tl{m}_o^2 \sim 0.1 \tl{m}_o^2|_0$, and so the quasi-fixed behaviour 
does not predict the squark mixing, it merely says that it will 
be fairly small. Thus information about the high-energy physics is retained in
the squark sector in the form of the mixings.

Finally, we stress that the improvement in our knowledge of the 
mass spectrum is due to the existence of {\em non-zero} \/
true fixed points towards which the soft terms are focusing 
in the infra-red, and is not simply due to the diverging 
gluino mass; if all the fixed points had been zero, the 
squark masses would all have been focused towards zero
and, relative to the average squark mass, would have been no 
better determined than at the GUT scale. 

\section{Discussion}
We now list the assumptions it is necessary to take in order for our
analytic solutions in Eq.(\ref{solutions}) to be accurate:
\begin{itemize}
\item{A desert consisting of the MSSM between $M_{SUSY}$ and $M_{GUT}$}
\item{Low $\tan \beta < 30$, such that $h_b$ and $h_\tau$ may be regarded as a
small perturbation}
\item{No {\em large} \/hierarchies in the soft masses}
\end{itemize}
The last assumption is natural in models of SUSY breaking which have no
flavour dependence. 
A violation of the last assumption may still not be enough to destroy the
validity of all of the QFP predictions~\cite{ourstobe}.

It is striking that 
just by increasing $h_t$ we greatly enhance the 
predictivity of the MSSM especially in the mass spectrum. 
Moreover, the soft mass predictions we find are broadly in agreement 
with electroweak symmetry breaking, and with phenomenology; the $m_2^2
$ term has a negative fixed point to which it is attracted, and in the
CMSSM the remaining squark mass-squareds are guaranteed to be positive
irrespective of their value at the GUT scale.
Perhaps the most interesting and powerful additional constraints may 
come from the avoidance of charge and colour breaking minima and potentials
which are unbounded from 
below~\cite{frere,komatsu,casas}. 
In the CMSSM a strong condition comes from the slepton
mass squareds~\cite{komatsu,casas};
\be
m_2^2+m_L^2 > 0.
\ee
From our solutions above this translates into $\tl{m}_1^2 > 0.72 $, which, 
taken at face value, means quite a strong constraint on $m_0$, 
\be
m_0^2 > 5.5 m_{1/2}^2
\ee
Whether this and all the other constraints are valid and can be
satisfied within the CMSSM will be investigated elsewhere. 

Another obvious area of relevance is flavour changing neutral current
(FCNC) phenomenology. The focusing behaviour tends to decrease the
FCNCs generated from splittings of GUT scale soft masses, supporting previous
numerical studies~\cite{pok}.
It is clear however, that the soft parameters do not focus enough  
to solve the notorious flavour problem associated with
the most general supersymmetric breaking. 
In fact, avoiding too large FCNCs typically implies a constraint 
on parameters such as  
\[
\frac{m^2_{\tl{s}}-m^2_{\tl{d}}} 
     {m^2_{\tl{s}}+m^2_{\tl{d}}}.
\]
These are not reduced by the orders of magnitude 
which would be required for a more general (non-universal) case. 

On the other hand we can say something about a 
picture of CP violation in the MSSM which was proposed recently 
in Ref.\cite{me2}, where it was suggested that the CP violation
observed in the Kaon system could be solely due to CP violation
in the $A$-terms of the third generation, with the CKM matrix 
being entirely real. In the quasi-fixed regime, we see that this scenario
cannot work when the top Yukawa is at the QFP or very large; the  
third generation $A$-terms run to QFPs which 
are independent of the initial conditions and which do not 
violate CP\@. As we have seen 
this is a quite general property of the MSSM\@. Thus, deep in the 
quasi-fixed regime of the MSSM, the only possible source of 
the observed CP violation is the CKM matrix. 

There are many other theoretical and phenomenological 
aspects of phenomenology where 
quasi-fixed behaviour could be of relevance. 
It would be interesting, for
example, to calculate what FCNCs the spectrum above predicts. 
A separate question is whether it is possible to provide similar 
(but more complicated) analytic solutions for the large $\tan \beta$ regime.

The inverse (`bottom up') approach to that taken in this paper 
was tried in Ref.\cite{fox}, where constraints were
placed upon GUT scale values from low energy FCNC constraints. 
However if, as seems likely, $h_t$ is far above its true-fixed point, 
then the `bottom up' approach fails (as the authors of  
Ref.\cite{fox} pointed out themselves);
a small error in the low energy constraint (such as two loop effects) 
produces a large error in the GUT scale parameters.
This is because, in trying to extract
information at the GUT scale, one has to fight against the infra-red 
fixed point structure (i.e.\ there is no information accessible for
$A_t(M_{GUT})$ precisely because it has a QFP). 

In summary, we have presented analytic solutions to the RGEs of the soft
parameters of the MSSM in the low $\tan \beta$ regime, 
and examined their quasi-fixed behaviour. 
They are valid when
there are no large hierarchies in the soft parameters. The solutions add to
our understanding of how the RGEs in the MSSM act.
They exhibit a strong focusing effect: i.e.\ the low energy
parameters are insensitive to their high energy values thus showing 
quasi-fixed behaviour. 
The $A_{U_{\alpha 3}}$ and $A_{U_{3 \alpha}}$ 
parameters actually have QFPs 
irrespective of the pattern of supersymmetry breaking, and in the 
CMSSM there exist other independent QFPs as well. For values 
of $h_t(M_{GUT}) \geqsim 1.1 $ quasi-fixed behaviour is a dominant
feature of the renormalisation group evolution in the MSSM\@. 

\section{Acknowledgments}
\noindent BCA would like to thank the ISN for support, and the ULB 
for hospitality extended whilst part of this work was being carried out.

\newpage

\begin{thebibliography}{99}

\bibitem{irfps}
B.~Pendleton and G.G.~Ross, Phys. Lett. {\bf B98} (1981) 21;
C.T.~Hill, Phys. Rev. {\bf D24} (1981) 691; B.C.~Allanach and S.F.~King,
RAL-TR-97-019, hep-ph/9705206; B.C.~Allanach, G.~Amelino-Camelia,
O.~Philipsen, Phys. Lett. {\bf B393} (1997) 349;
B.C.~Allanach and S.F.~King, RAL-97-007, SHEP-97-01, hep-ph/9703293.

\bibitem{graham}
M.~Lanzagorta and G.G.~Ross, Phys. Lett. {\bf B349} (1995) 31;
M.~Lanzagorta and G.G.~Ross, Phys. Lett. {\bf B364} (1995) 163.

\bibitem{largetanb}
B.~Schrempp, M.~Wimmer, hep-ph/9606386;
B.~Schrempp, Phys. Lett. {\bf B344} (1995) 193.

\bibitem{carena}
M.~Carena, M.~Olechowski, S.~Pokorski and C.~E.~M.~Wagner, 
Nucl. Phys. {\bf B419} (1994) 213; {\bf 426} (1994) 269;
M.~Carena and C.~E.~M.~Wagner, Nucl. Phys. {\bf B452} (1995) 45

\bibitem{casas}
J.~A.~Casas, A.~Lleyda, C.~Munoz, Phys.~Lett. {\bf B389} (1996) 305 

\bibitem{haberkane}
H.E.~Haber and G.L.~Kane, Phys. Rept. {\bf 117} (1985) 75.

\bibitem{me}
S.A.~Abel, W.N.~Cottingham, I.B.~Whittingham, Phys. Lett. {\bf B370} (1996)
106.

\bibitem{me2}
S.A.~Abel, J.M.~Fr\`ere, Phys. Rev. {\bf D55} (1997) 1623.

\bibitem{thresh}
P.~Langacker and N.~Polonsky,  Phys. Rev. {\bf D52} (1995) 3081.

\bibitem{rges}
S.P.~Martin and M.T.~Vaughn, Phys. Rev. {\bf D 50} (1994) 2282.

\bibitem{ourstobe}
S.A.~Abel and B.C.~Allanach, work in progress.

\bibitem{brax}
P.~Brax and C.~A.~Savoy, \npb{447}{227}{1995} 

\bibitem{barger}
V.~Barger, M.S.~Berger and P.~Ohmann, \prd{49}{4908}{1994};
G.L.~Kane, C.~Kolda, L.~Roszkowski {\em et al}, \prd{49}{6173}{1994}.

\bibitem{frere}
H.-P.~Nilles, M.~Srednicki and D.~Wyler, 
\plb{120}{346}{1983};
J.-M.~Fr\`ere, D.R.T.~Jones and S.~Raby, \npb{222}{11}{1983};
J.-P.~Derendinger and C.A.~Savoy, \npb{237}{307}{1984};
C.~Kounnas, A.B.~Lahanas, D.V.~Nanopoulos and M.~Quir\'os, 
\npb{236}{438}{1984};
M.~Drees, M.~Gl\"uck and K.~Grassie, \plb{157}{164}{1985};
J.F.~Gunion, H.E.~Haber and M.~Sher, \npb{306}{1}{1988};
P.~Langacker and N.~Polonsky, \prd{50}{2199}{1994}; 
J.A.~Casas and S.~Dimopoulos, \plb{387}{107}{1996};
J.A.~Casas, A.~Lleyda and C.~Mu\~noz, \plb{389}{305}{1996}.

\bibitem{komatsu}
H.~Komatsu, \plb{215}{323}{1988}.

\bibitem{pok}
D.~Choudhury, F.~Eberlein, A.~Konig, J.~Louis, 
S.~Pokorski, \plb{342}{180}{1995}.

\bibitem{fox}
M.~Carena, P.~Chankowski, M.~Olechowski, S.~Pokorski and
C.E.M.~Wagner, \npb{491}{103}{1997}


\end{thebibliography}
\end{document}